\documentclass{article}
\usepackage[T1]{fontenc} 
\usepackage[utf8]{inputenc} 
\usepackage{ismir,amsmath,cite,url}
\usepackage{graphicx}
\usepackage{color}
\usepackage{quoting}
\def\lbd{}	
\usepackage{booktabs}

\title{The Biased Journey of \texttt{MSD\_Audio.zip}}

\multauthor
{Haven Kim $^1$ \hspace{1cm} Keunwoo Choi $^2$
\hspace{1cm}Mateusz Modrzejewski $^3$ \hspace{1cm} Cynthia C. S. Liem $^4$}
{
  $^1$ Graduate School of Culture Technology, KAIST, South Korea\\
$^2$  Gaudio Lab, Inc., Seoul, South Korea / Prescient Design, New York, USA\\
$^3$  Institute of Computer Science, Warsaw University of Technology, Poland\\\
$^4$  Multimedia Computing Group, Delft University of Technology, The Netherlands\\
{\tt\small khaven@kaist.ac.kr, keunwoo@gaudiolab.com, mateusz.modrzejewski@pw.edu.pl, c.c.s.liem@tudelft.nl}
}

\def\authorname{Haven Kim, Keunwoo Choi, Mateusz Modrzejewski, and Cynthia C. S. Liem}

\begin{document}

\maketitle
\begin{abstract}
The equitable distribution of academic data is crucial for ensuring equal research opportunities, and ultimately further progress. Yet, due to the complexity of using the API for audio data that corresponds to the Million Song Dataset along with its misreporting (before 2016) and the discontinuation of this API (after 2016), access to this data has become restricted to those within certain affiliations that are connected peer-to-peer. In this paper, we delve into this issue, drawing insights from the experiences of 22 individuals who either attempted to access the data or played a role in its creation. With this, we hope to initiate more critical dialogue and more thoughtful consideration with regard to access privilege in the MIR community.
\end{abstract}

\section{Introduction}\label{sec:introduction}

\vspace{0.3cm}
\hspace{0.5cm}\textit{``It was on the school server."}
\vspace{-0.05cm}
\begin{center}
     \textit{vs.}
\end{center}
\vspace{-0.05cm}

\hspace{2.4cm}\textit{``We did not know whom to ask."}
\vspace{0.6cm}

The Million Song Dataset (MSD)~\cite{msd} has been a cornerstone for audio-centric music information retrieval (MIR) studies, such as music auto-tagging, with its significance widely acknowledged. However, access to audio data for this dataset (MSD Audio) is limited to peer-to-peer sharing since 2016, making it difficult to regard it as publicly available. As we will show, this limitation has led to disparities disadvantaging those affiliated with institutions that are personally less-connected within the MIR community, either geographically or academically. This has jeopardized the principle of equality within the community, as well as the reproducibility and advancement of previous research~\cite{fecher2015drives}.

In this paper, we will address this issue based on anecdotal comments from one individual who contributed to the creation of the dataset, as well as 21 individuals who have attempted to access the MSD audio. We collected these comments in two ways. First, we distributed a survey via the ISMIR mailing list, and those familiar with the dataset voluntarily participated. Second, after identifying approximately sixty papers that utilized the MSD audio for their experiments, we personally contacted the authors of each paper and invited them to either complete the survey or participate in an informal interview.

\section{From misreported public availability to institutional divide}

Initially designed as a large publicly available dataset, the MSD contains metadata for a million contemporary popular music tracks~\cite{msd}. The contributors have invested in aligning its metadata with another music audio set and API in order to facilitate various research opportunities that benefit from datasets. Unexpectedly, some researchers have used this capability to obtain 30-second audio previews of music tracks, which were matched with the MSD data by leveraging the API from ~\url{7digital.com}.



\begin{table}[]
\centering
\resizebox{\linewidth}{!}{%
\begin{tabular}{@{}cccc@{}}
\toprule
\textbf{Name} & \textbf{Country} & \textbf{When} & \textbf{How} \\ \midrule
TU Wien & Austria & 2011 & 7Digital API \\
Ghent University & Belgium & 2011-2013 & 7Digital API \\
NYU & United States & 2011-2013 & Peer-to-peer sharing \\
Columbia University & United States & 2014 & Peer-to-peer sharing \\
University of Oxford & United Kingdom & 2014-2015 & Peer-to-peer sharing \\
University of Edinburgh & United Kingdom & 2014-2015 & Peer-to-peer sharing \\
UPF & Spain & 2014-2015 & Peer-to-peer sharing \\
Academia Sinica & Taiwan & 2015 & Peer-to-peer sharing \\
QMUL & United Kingdom & 2015 & Peer-to-peer sharing \\
Johns Hopkins University & United States & 2016 & Peer-to-peer sharing \\
TU Delft & Netherlands & 2016 & Peer-to-peer sharing \\
KAIST & South Korea & 2016 & Peer-to-peer sharing \\
Deezer & France & 2011-2016 & U/I \\
JKU & Austria & U/I & U/I \\ \bottomrule
\end{tabular}
}
\caption{Organizations that have accessed MSD Audio by 2016, including the approximate years and methods of access. U/I indicates unidentified.}
\label{tab:list}
\end{table}

Based on anecdotes from researchers who attempted to access audio previews using the API, comprehensively scraping audio previews was nearly impossible without significant financial resources or technical expertise. We assume only two organizations succeeded in this
before the deactivation of the \url{7digital.com} API. In fact, most of the organizations that we identified as having access to this data (listed in Table~\ref{tab:list}) acquired it through direct and informal peer-to-peer sharing.
Those who obtained the data through direct sharing were, naturally, those connected peer-to-peer with data-owning organizations. MSD Audio is prohibitively large (about 700 GB) to easily share through online transfer. As a result, geographical proximity seems to have played a role.

Yet, none of those who acquired the data through peer-to-peer sharing and utilized it publicly acknowledged the practical unavailability of this dataset. We received comments that those who acquired the dataset through peer-to-peer sharing reported in their papers that they had obtained the data via web scraping. This misinformation further confused researchers outside of major organizations within the MIR community, leading to numerous unsuccessful attempts to obtain data by web scraping. This unequal accessibility to MSD Audio has widened the gap between majorities and minorities within the community. 

\section{Anecdotal Analysis : Unequal Accessibility}

In this section, we will analyze anecdotal comments from 21 individuals who have tried accessing MSD Audio, which we collected through surveys or interviews, where we asked about the methods, results, and the approximate timings of their attempts, as well as their affiliations and professions at the time of their attempts, to address the issue of unequal accessibility.

Peer-to-peer sharing instances that we identified primarily have occurred between organizations that owned the data and those closely connected to them, either geographically or academically, with the most recent instance of peer-to-peer sharing occurring in 2016. All individuals who attempted to obtain the dataset in 2017 or later and succeeded (5 individuals) were affiliated with organizations that owned the data. Conversely, all those from organizations without the data who tried in 2017 or later (6 individuals) experienced at least one unsuccessful attempt, where two of these individuals ultimately abandoned the research project they had initially planned. Four of those who were unsuccessful mentioned that they had even tried asking individuals outside their organizations, but to no avail. One of them specifically noted that they currently lack access to the dataset because they ``do not know whom to ask''. This institutional divide seems to lead to inequality among industrial organizations as well. One respondent evidenced this by reporting that they had no access to this data while employed at an organization with less than 50 employees, but immediately obtained access upon moving to an organization with more than 500 employees. 

Our further investigation into the unsuccessful attempts supports the existence of institutional divides along with geographical and prestige bias. Of the 7 unsuccessful peer-to-peer sharing attempts we identified from the anecdotes we collected, five came from relatively less active organizations within the MIR community, without first- or last-author papers at the ISMIR conference in the past three years. The remaining two were non-Western organizations. On the other hand, all of the organizations we identified to own access to this data are notably either prominent within the MIR community, with at least five papers accepted at the ISMIR conference in the past three years, or Western-based, as provided in Table~\ref{tab:list}.

In addition to organization prestige, individual research experience also appears to influence the success of requests, favoring experienced researchers over novices. For instance, we observed two separate attempts made by those from the same organization, where an undergraduate student was unable to obtain the data despite the request, whereas a faculty member's request was successful.

\section{Conclusions}
In this paper, we delved into the unequal accessibility issue concerning MSD audio, which divides the MIR community into those who can access the data and those who cannot. This problem arises not only due to the complexity and discontinuation of using APIs, but also because of information provided by authors who claimed to have acquired the data through web scraping when, in fact, they obtained it via peer-to-peer sharing. This has disproportionately affected researchers who are not closely connected to the data owning organizations and those with limited research experience, sidelining minorities within the MIR community.

Our data of those who had access to MSD Audio is quite comprehensive, as there are few papers that are based on the dataset in the early years. However, the collected failure cases are far from being comprehensive; our survey is distributed via the ISMIR mailing list or personal contacts to authors who utilized the data for their papers, 
which already set a strong survivorship bias. One of the authors has received many inquiries about MSD Audio so far, but none of them participated in our survey. It is indeed not possible to discover all the hidden, doomed attempts towards MSD Audio.

This situation, ever since 2011, challenges us to imagine how much MIR research could have been done if MSD Audio was available more widely and equally, and how many potential MIR researchers could have had a more active and successful research career. Yet we have been overlooking this problem, rather than facing it, because it is only a severe problem to those who do not have a voice in the research community. By presenting this survey and these anecdotes, we advocate for more inclusive and transparent data accessibility and research  opportunities, and hope to cultivate a more diverse, equitable, and productive MIR research landscape.

\bibliography{ISMIRtemplate}

\begin{thebibliography}{1}
\providecommand{\url}[1]{#1}
\csname url@samestyle\endcsname
\providecommand{\newblock}{\relax}
\providecommand{\bibinfo}[2]{#2}
\providecommand{\BIBentrySTDinterwordspacing}{\spaceskip=0pt\relax}
\providecommand{\BIBentryALTinterwordstretchfactor}{4}
\providecommand{\BIBentryALTinterwordspacing}{\spaceskip=\fontdimen2\font plus
\BIBentryALTinterwordstretchfactor\fontdimen3\font minus
  \fontdimen4\font\relax}
\providecommand{\BIBforeignlanguage}[2]{{%
\expandafter\ifx\csname l@#1\endcsname\relax
\typeout{** WARNING: IEEEtran.bst: No hyphenation pattern has been}%
\typeout{** loaded for the language `#1'. Using the pattern for}%
\typeout{** the default language instead.}%
\else
\language=\csname l@#1\endcsname
\fi
#2}}
\providecommand{\BIBdecl}{\relax}
\BIBdecl

\bibitem{msd}
T.~Bertin-Mahieux, D.~P. Ellis, B.~Whitman, and P.~Lamere, ``The million song
  dataset,'' in \emph{12th International Society for Music Information
  Retrieval Conference}, 2011, pp. 591--596.

\bibitem{fecher2015drives}
B.~Fecher, S.~Friesike, and M.~Hebing, ``What drives academic data sharing?''
  \emph{PloS one}, vol.~10, no.~2, p. e0118053, 2015.

\end{thebibliography}

%
%
%
%
%

\end{document}